\documentclass[journal=apchd5, manuscript=article]{achemso}

\usepackage{wrapfig}
\usepackage{float}
\usepackage{amssymb}
\usepackage{amsmath}
\usepackage{graphicx}
\usepackage{dcolumn}
\usepackage{bm}
\usepackage{amsmath}
\usepackage{soul}
\usepackage{xcolor}
\usepackage[dvipsnames]{xcolor}
\usepackage[colorlinks=true,allcolors=blue]{hyperref}

\date{\today}
\title{Photonic-Crystal Microresonator Frequency Combs in the O-band}

\author{Sarang Yeola}
\affiliation{Time and Frequency Division, National Institute of Standards and Technology, Boulder, Colorado 80305, USA}
\alsoaffiliation{Department of Physics, University of Colorado Boulder, Boulder, Colorado 80309, USA}

\author{Alexa R. Carollo}
\affiliation{Time and Frequency Division, National Institute of Standards and Technology, Boulder, Colorado 80305, USA}

\author{Jizhao Zang}
\affiliation{Time and Frequency Division, National Institute of Standards and Technology, Boulder, Colorado 80305, USA}
\alsoaffiliation{Department of Electrical Engineering, University of Colorado Denver, Denver, Colorado 80204, USA}

\author{Scott B. Papp}
\affiliation{Time and Frequency Division, National Institute of Standards and Technology, Boulder, Colorado 80305, USA}
\alsoaffiliation{Department of Physics, University of Colorado Boulder, Boulder, Colorado 80309, USA}
\email{scott.papp@nist.gov}

\begin{document}

\maketitle  

\begin{abstract}

Photonic-crystal microresonators (PhCRs) are a powerful platform for generating Kerr frequency combs. Because Kerr-soliton dynamics in PhCRs are largely decoupled from the operating wavelength, the comb output can be engineered through customization of the device layer. Here, we demonstrate a tantalum pentoxide (tantala) PhCR platform that supports 1310 nm and 1550 nm band operation, and we explore high-efficiency O-band soliton microcombs with all-semiconductor laser pumps. We engineer the PhCRs with silicon dioxide cladding and normal dispersion with intrinsic quality factors exceeding $7\times10^{6}$. By pumping bandgap modes, we obtain robust and efficient soliton comb formation at a 200 GHz mode spacing. Our PhCRs enable systematic tuning from narrowband to broadband comb states within a single device geometry. The combs exhibit low relative intensity noise approaching the shot-noise limit, indicating stable phase-matching in the PhCR. Using a second resonator coupler, we amplify the comb output off-chip, demonstrating a pathway to high-power O-band sources. These results establish PhCR engineering in the tantala platform as a scalable approach to wavelength-agile, low-noise microcombs for applications in communications, sensing, and signaling.

\end{abstract}

\section{Introduction}

The recent growth of integrated-photonics technologies provides routes to compact and efficient multi-wavelength light sources at the $\approx200 $ GHz mode spacing optimized for low power consumption data links \cite{Mehta_2023, Song_2026, pirmoradi_integrated_2025, yu_continuum_2022, zang_laser_2025, jinBandgapdetunedExcitationRegime2025, helgason_surpassing_2023,rizzo_massively_2023, pirmoradi_single_2025, omirzakhov_single_2025, rakowski_45nm_2020}. Yet there is a growing need to translate and connect these technologies across different spectral bands. The O-band, spanning 1260 nm to 1360 nm, is widely used in optical communications because standard single-mode fiber exhibits near-zero chromatic dispersion in this wavelength range \cite{belykh_o-band_2026}. Together with mature and efficient semiconductor laser technologies near 1310 nm, these properties enable simple and cost-effective short-reach optical links. Furthermore, in biological applications, the 1300 nm window enables deeper tissue penetration and reduced scattering for techniques like optical coherence tomography \cite{kodach_quantitative_2010}. The O-band, however, currently lacks the infrastructure of the C-band, particularly regarding high-gain solid-state amplifiers and low loss integrated nonlinear photonics.  

There are several types of laser technologies that are relevant for O-band applications. Distributed feedback (DFB) and Bragg reflector semiconductor lasers are widely used in high-speed telecommunications because of their frequency stability and narrow linewidth \cite{belykh_o-band_2026, malik_widely_2020, han_electrically_2021}. 
For applications requiring higher power, solid-state bulk lasers, such as chromium-doped forsterite \cite{mckinnieChromiumdopedForsteriteInfluence1996a} and neodymium-doped crystals \cite{bethea_megawatt_1973, lu_simultaneous_2013}, provide an alternative capable of generating femtosecond pulses and high peak intensities, though they are often limited by their reliance on precise free-space optical alignment and susceptibility to thermal lensing. More flexible systems include optical fiber lasers and amplifiers, such as praseodymium-doped \cite{ohishiPr3+dopedFluorideFiber1991} and bismuth-doped fiber systems \cite{donodinBismuthDopedFiberAmplifiers2024}, which provide high gain and fiber compatibility. At the highest powers are iodine gas lasers, which can achieve megawatt-scale output powers near 1315~nm, though they generally have a large footprint and are resource intensive \cite{barmashenkoChemicalLasersCOIL2021}.

Kerr microresonator solitons are versatile frequency comb sources \cite{spencer_optical-frequency_2018}, and recent progress has demonstrated that stable generation of O-band soliton microcombs is possible through techniques such as pump-phase modulation \cite{dingOpticalObandSoliton2023} and laser self-injection locking \cite{jiOBandPlaticonMicrocomb2026}, which stabilize soliton and platicon formation. In addition, bismuth-doped fiber amplifiers have been employed with microcombs to provide flat gain across the O-band wavelength region \cite{stoliarovUnlockingOBandHighpower2026}. Photonic-crystal resonators (PhCRs) provide a powerful platform for refining the characteristics and operating envelope of soliton microcombs \cite{yu_spontaneous_2021,liuImplementingPhotoniccrystalResonator2025, liu_next-generation_2025,zang_foundry_2024}. These devices facilitate universal phase matching for four-wave mixing and soliton generation, enabling precise spectral control without sacrificing key microcomb properties. Furthermore, the PhCR architecture provides direct control over critical metrics, including frequency spacing, conversion efficiency, and spectral bandwidth.

\begin{figure*}[t!]
    \centering
    \includegraphics[width=\textwidth]{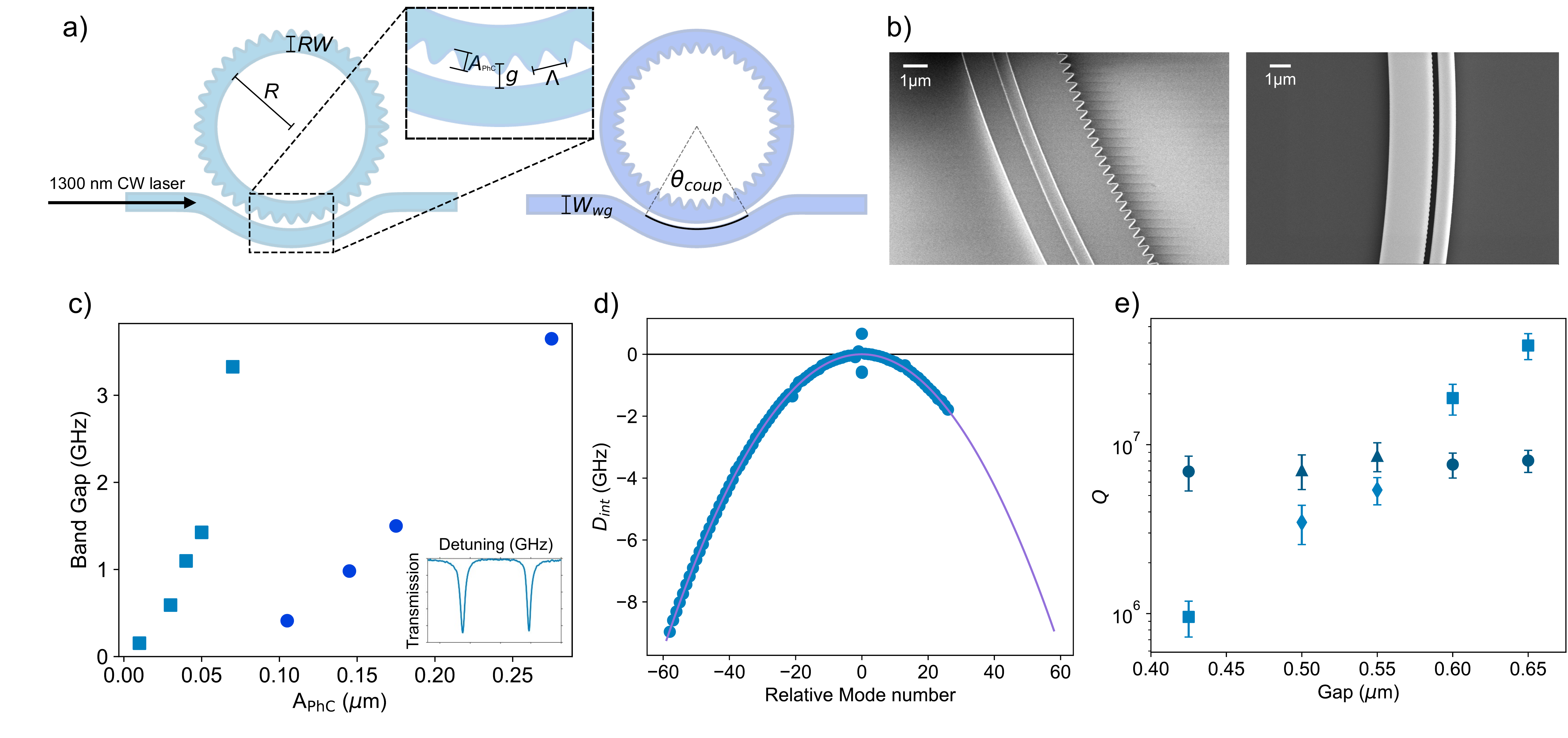}
    \caption{Resonator design and fabrication. a) Resonator schematic with photonic crystal design parameters that include amplitude $A_\text{PhC}$, spacing $\Lambda$, and choice of inner or outer sidewall modulation. Bus waveguide parameters include waveguide width $W_{wg}$, gap between waveguide and resonator $g$, and coupling angle $\theta_{coup}$. b) Scanning electron microscope images of the photonic crystal resonator. c) Bandgap dependence on $A_\text{PhC}$ for outer (squares) and inner (circles) modulation. Band gaps are calculated from fits of the transmission traces (inset).  d) Integrated dispersion as a function of relative mode number. Circles indicate the measured modes, and the purple line shows the fitted $D_\text{int}$ to second order. e) Quality factor dependence on gap $g$. Circles and triangles are intrinsic quality factors $Q_i$, and squares and diamonds are coupling quality factors $Q_c$. Quality factors for each devices are averaged over modes in the O-band; error bars indicate one standard deviation. A difference in waveguide width in the devices tested is indicated by the choice of diamond/triangles and circle/squares.}
    \label{fig:design}
\end{figure*}

Here, we demonstrate O-band dark soliton microcombs in normal-dispersion PhCRs, extending bandgap-engineered Kerr comb generation to this technologically important wavelength range. We present the design and characterization of PhCRs in a fully oxide-clad titania–tantala platform, where this metal-oxide mixture enables high intrinsic quality factor, deterministic bandgap control, and precise dispersion engineering.  Our platform supports soliton microcomb generation from the O- band to the S-, C-, and L- bands by design of the device layer. Using devices for the O-band, we show photonic-crystal mode bandgaps provide a mechanism for microcomb spectral control, enabling tuning of center wavelength, bandwidth, and power distribution. We further demonstrate spontaneous soliton generation by scanning the pump laser into resonance, achieving stable operation with shot-noise-limited intensity noise. Finally, by operating PhCRs with integrated drop ports, we directly access and amplify the comb output, establishing compatibility with semiconductor optical amplifiers and enabling high per-mode power operation.

\section{Experimental Section}

To realize O-band microcombs, we carry out the design of PhCRs with the goal of normal group-velocity dispersion, 200 GHz free spectral range (FSR), and a photonic-crystal bandgap in the O-band. Here, the relevant design parameters that largely control the dispersion are the microresonator ring radius ($R$) and ring width ($RW$); see Fig. \ref{fig:design}a. Additionally, we design a photonic crystal as a periodic modulation to either the inner or the outer sidewall of the resonator, with spatial period $\Lambda = \pi R/m$ to generate a bandgap on the pump mode, $m$. The magnitude of the bandgap is related to the amplitude of the modulation, $A_{\rm{PhC}}$. We aim for bandgaps large enough to enable phase matching and high power across many comb lines. For control over the coupling efficiency, we design pulley couplers and vary the waveguide width, gap, and coupling angle of the bus waveguide. We use simulations of microresonator dispersion to guide our designs in the titania-tantala platform \cite{Moille:19, zang_laser_2025} .

\begin{figure*}[t]
    \centering
    \includegraphics[width = 0.8 \textwidth]{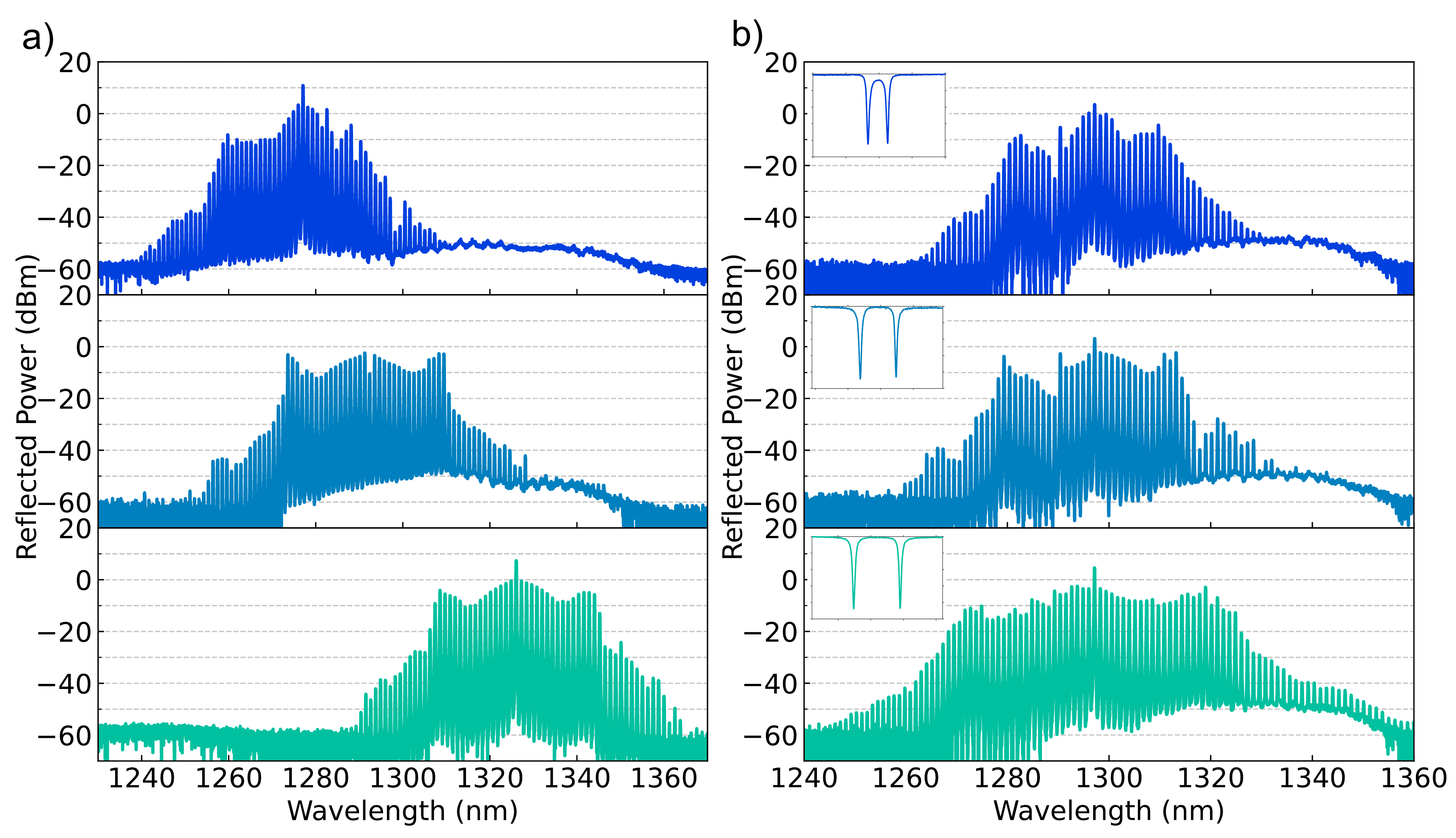}
    \caption{a) Comb spectra generated by pumping devices with photonic crystal bandgaps designed at different wavelengths: (top) 1276.5 nm, (middle) 1291.2 nm, (bottom) 1326.5 nm. b) Comb spectra generated by pumping devices with the same bandgap wavelength, but with different bandgap sizes: (top) 0.59 GHz, (middle) 1.1 GHz, (bottom) 1.43 GHz. Insets show the relative bandgap sizes across the devices.}
    \label{fig:combs}
\end{figure*}

We fabricate PhCRs designed for the O-band, using the titania-tantala (TiO$_2$:Ta$_2$O$_5$) metal-oxide mixture \cite{carolloAmorphousMetalOxide2026,brodnik_monolithic_2026}. With ion-beam sputtering (IBS) at room temperature, we deposit the film mixture by co-sputtering titanium and tantalum metal targets. We grow 625 nm thick films on the SiO$_2$ layer of a thermally oxidized silicon wafer. We use electron beam lithography and an alumina hard mask to transfer the designs, and etch the films into PhCRs using fluorine-based reactive ion etching. We achieve nanoscale features necessary for inducing a photonic bandgap on a selected azimuthal mode, as shown in the SEM images in Fig \ref{fig:design}b. From SEM images of the coupling region between the bus waveguide and resonator, we verify that the sidewall corrugations---on both the inner (left) and outer (right) sidewall---are present. As shown, the corrugation on the outer sidewall is small, due to its smaller $A_{\rm{PhC}}$ design than the device with the inner sidewall corrugation. For optimal device performance, we anneal the wafer in air at 500 $^{\circ}$C to reduce the defect density in the titania-tantala film layer. To protect the devices, we deposit oxide cladding via inductively-coupled plasma-enhanced chemical vapor deposition. 

We test PhCR device bandgap, dispersion, and quality factors by scanning a tunable laser across the 1260 nm to 1360 nm wavelength range and measuring the transmission through the resonator.
We observe resonances as dips in the transmission spectrum and fit each resonance to a Gorodetsky model \cite{gorodetskyRayleighScatteringHighQ2000} to extract the resonance frequency, quality factors, and bandgap size. The photonic crystal splits the pump mode, with a bandgap magnitude that increases with $A_\text{PhC}$; see Fig. \ref{fig:design}c. The inset, which is a transmission trace around the pump resonance mode, shows the induced bandgap. Here, we explore how the location of the sidewall modulation affects the bandgap magnitude. An outer sidewall modulation requires a smaller $A_\text{PhC}$ than an inner modulation to achieve the same bandgap. This splitting affects the dispersion of the device, which is shown in the integrated dispersion plot, $D_\text{int}$, in Fig \ref{fig:design}d. A polynomial fit of the mode numbers $\mu$ indexed around the pump mode and the corresponding resonance frequencies $\omega_\mu$ is used to determine the Free spectral range (FSR) and second-order dipsersion term ($D_2$). The $D_\text{int}$ is calculated from the equation
\begin{equation}
    D_{\text{int}}(\mu) = \omega_\mu - (\omega_0 + (FSR) \mu) = D_2 \mu^2 / 2,
\end{equation}
with experimentally determined points shown as blue circles and a quadratic fit shown as a purple line. Here, the split mode is plotted as two separate modes with $\mu=0$.
With 625 nm thickness, 106 $\mu$m ring radius, and 3 $\mu$m to 4 $\mu$m ring width, the devices in this paper exhibit normal dispersion with $D_2$ of $-5$ MHz to $-6$ MHz at FSR of approximately 200 GHz. By pumping the lower-frequency pump mode, we enable phase matching for solitons in PhCRs with normal dispersion \cite{jinBandgapdetunedExcitationRegime2025}.

The extracted quality factors from the Gorodetsky model, including the intrinsic quality factor $Q_i$ and coupling quality factor $Q_c$, for each mode are averaged over the scan range to determine $Q$'s for each device. The resonators consistently exhibit high $Q_i$ over 7 million, as shown in Fig. \ref{fig:design}e. By adjusting the coupler parameters, we control $Q_c$, where $Q_c$ increases with gap. A smaller gap increases coupling to the waveguide, which we can adjust to support higher coupling efficiencies of pump light to the resonator. 

\section{Results}

\begin{figure*}[t!]
    \centering
    \includegraphics[width= 0.75\textwidth]{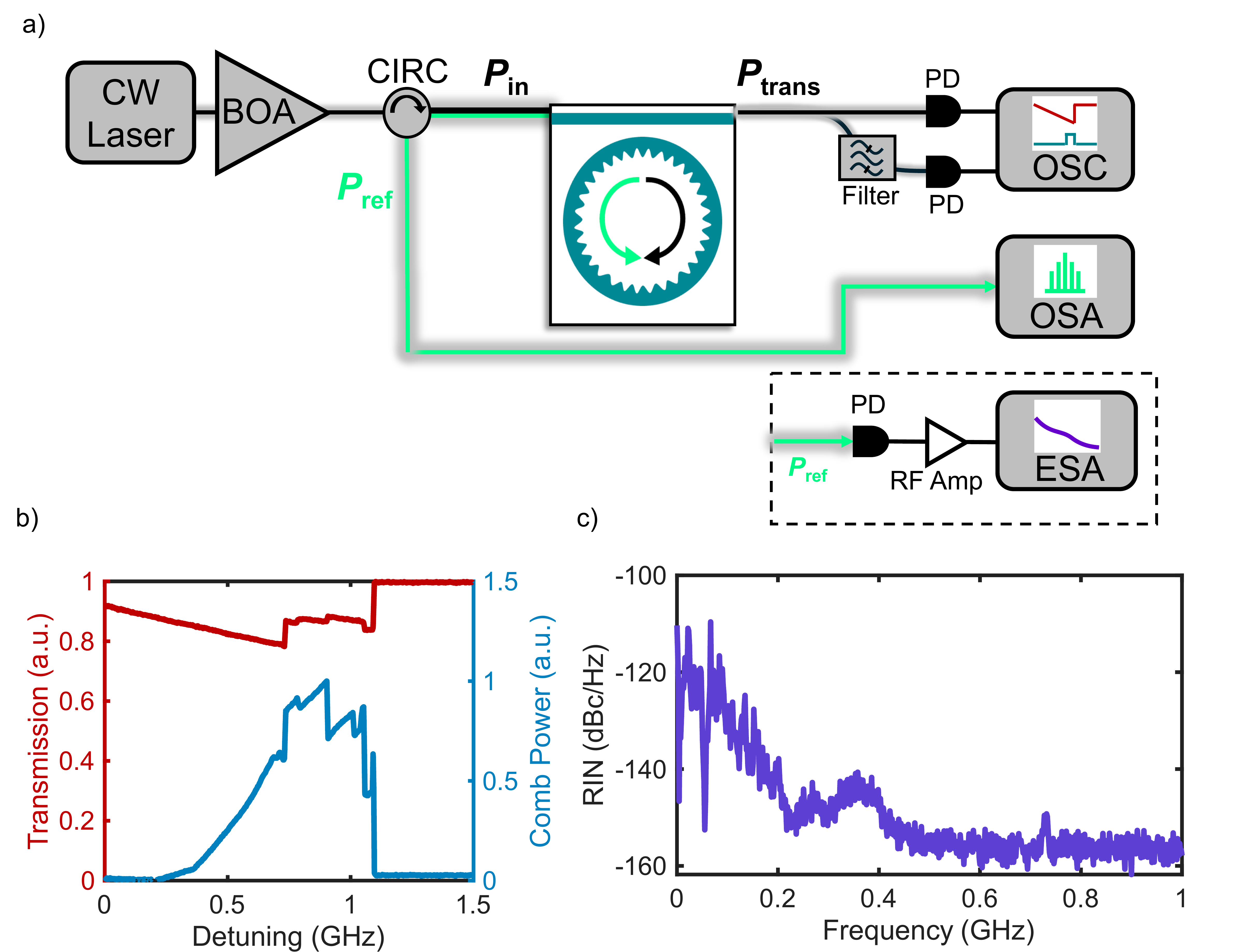}
    \caption{a) Experimental setup for generating combs in PhCRs, with inset showing the setup used for measuring their noise properties. CIRC: circulator; BOA: booster optical amplifier; PD: photodiode; OSC: oscilloscope; OSA: optical spectrum analyzer; RF Amp: radio-frequency amplifier; ESA: electrical spectrum analyzer. b) Transmission intensity (red) and comb intensity (blue) of a pumped microresonator as a function of detuning. c) Relative intensity noise spectrum of a comb.}
    \label{fig:rin}
\end{figure*}

To demonstrate control over O-band soliton microcomb formation in PhCRs, we explore devices with a range of design parameters; see Fig. \ref{fig:combs}. Specifically, we characterize how PhCR geometric design parameters affect spectral amplitude, width, and efficiency. By mapping these variables to the experimental spectra, we can establish a framework for tailoring microcomb properties.

The handful of tools provided by PhCR design, including bandgap wavelength and magnitude, allow us to control the properties of the generated combs. Particularly, by controlling the bandgap wavelength with $m$, we generate combs at varying pump wavelengths as shown in Fig. \ref{fig:combs}a. This provides access to the entire O-band by designing the PhCR mode for different pump wavelengths. Across these devices, we vary the coupling factor, $K = Q_i/Q_c$, with the goal of achieving high conversion efficiency. We define conversion efficiency as the percent of the pump power that is converted to comb power. For the device plotted in Fig. \ref{fig:combs}a (middle), with $K \approx 2.5$, we estimate a conversion efficiency near 20 \%. Furthermore, we can tune the spectral bandwidth by controlling the bandgap magnitude; see Fig. \ref{fig:combs}b. Pumping devices with larger $A_{\rm{PhC}}$ produces a series of combs with increasing spectral widths. At small bandgaps, most of the power in the resonator is concentrated to a few lines around the pump. As the bandgap increases, the phase-matching extends over a broader wavelength range and the comb power is distributed across more lines. These results show that comb width and power can be controlled by optimizing the PhCR design parameters, specifically the bandgap wavelength, magnitude, and waveguide coupling.

To obtain the spectra presented in Fig. \ref{fig:combs}, we characterize the devices using the experimental setup shown in Fig. \ref{fig:rin}a. 
We use a tunable continuous wave laser amplified by a booster optical amplifier (BOA) to pump the devices. 
The BOA is a semiconductor laser amplifier that operates in the O-band.
The input pump power, $P_\text{in}$, is coupled into the bus waveguide of the resonator using a lensed fiber.
The PhCR induces backscattering of the pump, thus we analyze both the transmitted ($P_\text{trans}$) and reflected ($P_\text{ref}$) outputs. 
Light propagating in the forward, or transmitted, direction is used to monitor the comb states as the pump detunes across the lower-frequency PhCR resonance mode; see Fig \ref{fig:rin}b. 
Here, we collect the transmitted output with a lensed fiber.
The transmission trace shows the depletion of the pump as the laser detuning increases.
By taking a portion of the signal and filtering out the residual pump, we observe the evolution of the comb power as a function of detuning, showing various comb states across the range.
The PhCRs typically generate most of the comb power in the backward, or reflected, direction.
We use an off-chip fiber circulator to separate the backwards propagating comb from the input pump.
To measure the spectra, we send the signal to an optical spectrum analyzer (OSA).
The comb powers in Fig. \ref{fig:combs} are calibrated to the on-chip power by adding back the power lost when coupling from the output facet to the lensed fiber (typically $\sim$2.5 dB/facet).

\begin{figure*}[ht]
    \centering
    \includegraphics[width = 0.75 \textwidth]{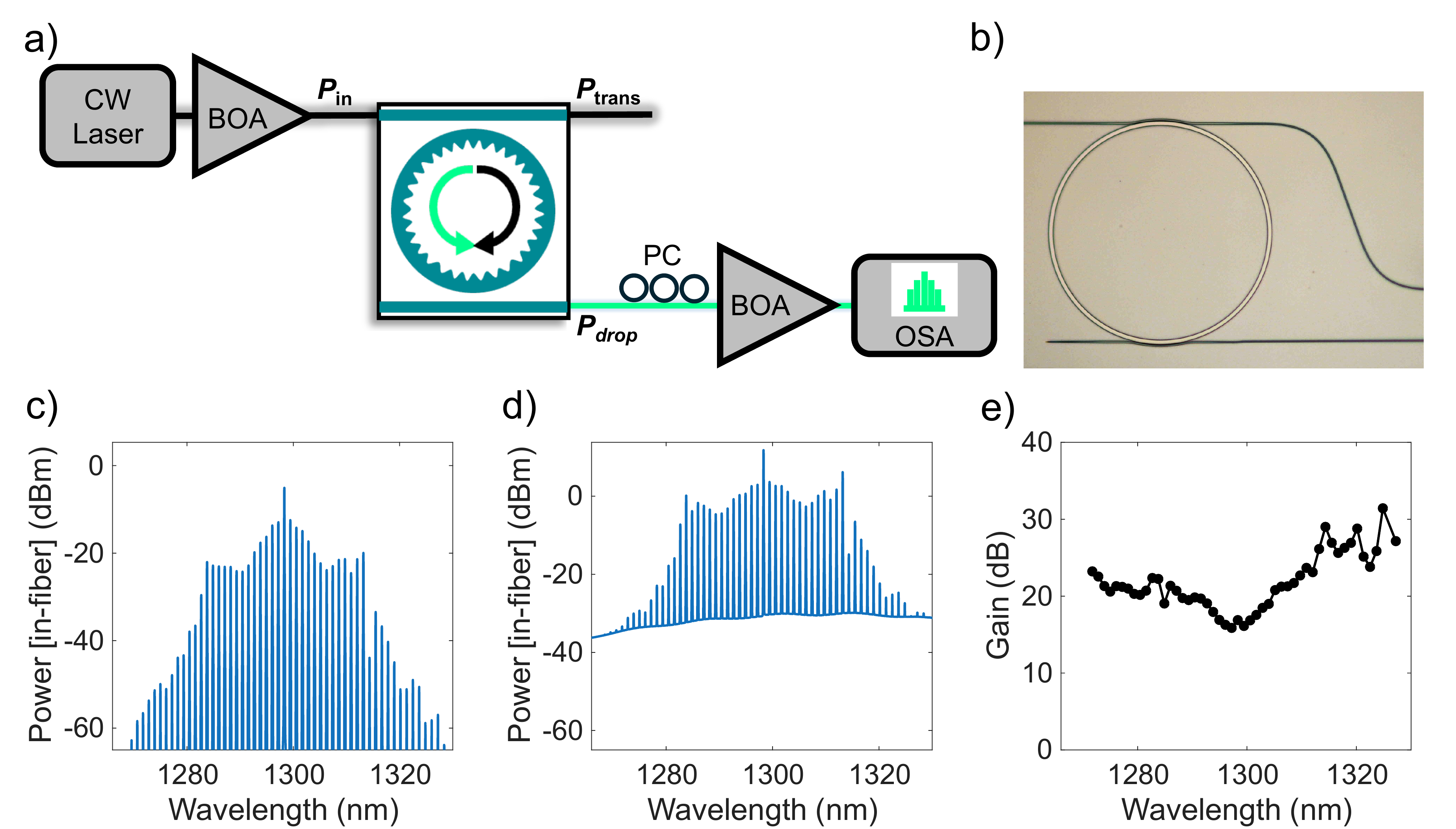}
    \caption{ a) Schematic of the drop port design for accessing the comb. PC: polarization control. b) Optical microscope image of a PhCR with a drop port. c) Comb spectrum measured at the drop port. d) Comb spectrum after amplification with a BOA. e) Gain measured per mode after amplification. }
    \label{fig:applications}
\end{figure*}

The relative intensity noise (RIN) of a comb for applications like data communication is a critical metric because it specifies the capacity of data each comb mode can support.
Here, we characterize the noise from the resonator's reflected port, as shown in the inset of Fig. \ref{fig:rin}a. 
To measure the noise, we send the comb into an amplified high-speed photodetector and measure the noise power spectral density with an electrical spectrum analyzer.
The noise power spectral density is normalized by the average DC power and the resolution bandwidth to determine the RIN in dBc/Hz.
Our combs typically have total on-chip powers between 1 mW and 10 mW.
For the comb represented in Fig \ref{fig:rin}c, at a photocurrent of 0.72 mA, the shot-noise floor is at -154 dBc/Hz.
The measured noise floor of the comb closely tracks the shot-noise limit, demonstrating that the system is not dominated by technical or thermal noise. 
This performance suggests that the comb modes are suitable as high-fidelity data-transmission carriers.

The realization of O-band soliton microcombs on the oxide-clad, titania-tantala platform provides a route to scalable integrated photonics and fully integrated systems. The oxide cladding serves to ensure environmental stability and provides mechanical isolation necessary for robust operation. However, the tendency for normal-dispersion PhCRs to generate combs in the backwards direction limits their uses to applications where off-chip circulators can be implemented. To facilitate practical uses and overcome this limitation, we explore the addition of a drop-port coupler; see Fig. \ref{fig:applications}a. We use this design to extract half of the comb power, which would otherwise be directed into the backward direction, into the forwards direction via the drop port. This is useful for applications requiring the benefits of the PhCR platform (e.g. high conversion efficiency, wavelength tunability), but where using an off-chip circulator is not possible, as with a fully integrated system. To explore this design, we fabricate PhCRs with a drop-port coupler; see Fig. \ref{fig:applications}b. Another advantage of using a drop port is that it filters out amplified spontaneous emission (ASE) from the laser; see Fig. \ref{fig:applications}c. Indeed, the broad ASE background we typically observe is absent in the combs measured from the drop port. Our results are consistent with prior work characterizing the effect that a microresonator drop port has on the ASE in ordinary devices without PhCRs \cite{wangDropportStudyMicroresonator2013}. This is advantageous because it lowers the noise floor.

Besides noise, another consideration for practical applications is meeting power requirements. One method for meeting power requirements is off-chip amplification. This method requires an amplifier with a broad gain bandwidth to be compatible with the spectral range of the comb. Since we validated that a BOA is a suitable pump source that can provide sufficient power to generate combs across the O-band, we explored implementing a second BOA to amplify the drop-port comb. To optimize the signal output of the polarization-sensitive amplifier, polarization control paddles were installed at the drop-port output. The resulting amplified comb (Fig. \ref{fig:applications}d) contains high-powered lines up to 5 mW per mode. This represents a gain of over 20 dB for several comb lines when using the external BOA; see Fig. \ref{fig:applications}e. This ability to generate high power combs is promising for future applications in optical data transmission.

\section{Conclusion}
In conclusion, we demonstrate O-band normal dispersion microcombs in PhCRs on an oxide-clad, titania-tantala integrated photonics platform. By adjusting device geometry, we control spectral properties including center wavelength and bandwidth, with the ability to generate signals with low noise. We also demonstrate integrated drop-port waveguides and achieve high-power-per-mode combs through semiconductor amplifiers, highlighting the versatility of the platform and its compatibility with existing technologies. These results provide a route to achieving low-noise O-band microcomb sources for potential applications in data communications and sensing.

\section{Acknowledgment}

This research has been funded by AFOSR FA9550-20-1-0004 Project Number 19RT1019, NSF Quantum Leap Challenge Institute Award OMA – 2016244, DARPA NaPSAC, and NIST. The authors declare no competing interests. This work is a contribution of the US Government and is not subject to US copyright. Mention of specific companies or trade names is for scientific communication only and does not constitute an endorsement by NIST. This document has not been peer reviewed but has been cleared by NIST for release.

\section{References}
\bibliography{main.bib}

\end{document}